\documentclass[runningheads]{llncs}
\usepackage{graphicx}
\usepackage{caption}
\usepackage{subcaption}
\usepackage{placeins}
\usepackage[utf8]{inputenc}
\usepackage{float}

\graphicspath{{Images/}}
\begin{document}
\title{PathRTM: Real-time prediction of KI-67 and tumor-infiltrated lymphocytes}
%
%\titlerunning{Abbreviated paper title}
% If the paper title is too long for the running head, you can set
% an abbreviated paper title here
%
\author{Steven (Zvi) Lapp\inst{1} \and
	Eli (Omid) David\inst{2} \and
	Nathan S. Netanyahu\inst{3}}
\authorrunning{S.Z Lapp, E.O. David, and N.S. Netanyahu}
% First names are abbreviated in the running head.
% If there are more than two authors, 'et al.' is used.
%
\institute{Department of Computer Science, Bar-Ilan University, Ramat-Gan 5290002, Israel 
	\email{lappzvi@gmail.com, mail@elidavid.com, nathan@cs.biu.ac.il}\\
}
\maketitle              % typeset the header of the contribution
\begin{abstract}
	In this paper, we introduce PathRTM, a novel deep neural network detector based on RTMDet, for automated KI-67 proliferation and tumor-infiltrated lymphocyte estimation. KI-67 proliferation and tumor-infiltrated lymphocyte estimation play a crucial role in cancer diagnosis and treatment. PathRTM is an extension of the PathoNet work, which uses single pixel keypoints for within each cell. We demonstrate that PathRTM, with higher-level supervision in the form of bounding box labels generated automatically from the keypoints using NuClick, can significantly improve KI-67 proliferation and tumor-infiltrated lymphocyte estimation. Experiments on our custom dataset show that PathRTM achieves state-of-the-art performance in KI-67 immunopositive, immunonegative, and lymphocyte detection, with an average precision (AP) of 41.3\%. Our results suggest that PathRTM is a promising approach for accurate KI-67 proliferation and tumor-infiltrated lymphocyte estimation, offering annotation efficiency, accurate predictive capabilities, and improved runtime. The method also enables estimation of cell sizes of interest, which was previously unavailable, through the bounding box predictions.
\end{abstract}
\section{Introduction}

\begin{figure}
    \centering
    \begin{subfigure}[b]{0.3\textwidth}
        \centering
        \includegraphics[width=\textwidth]{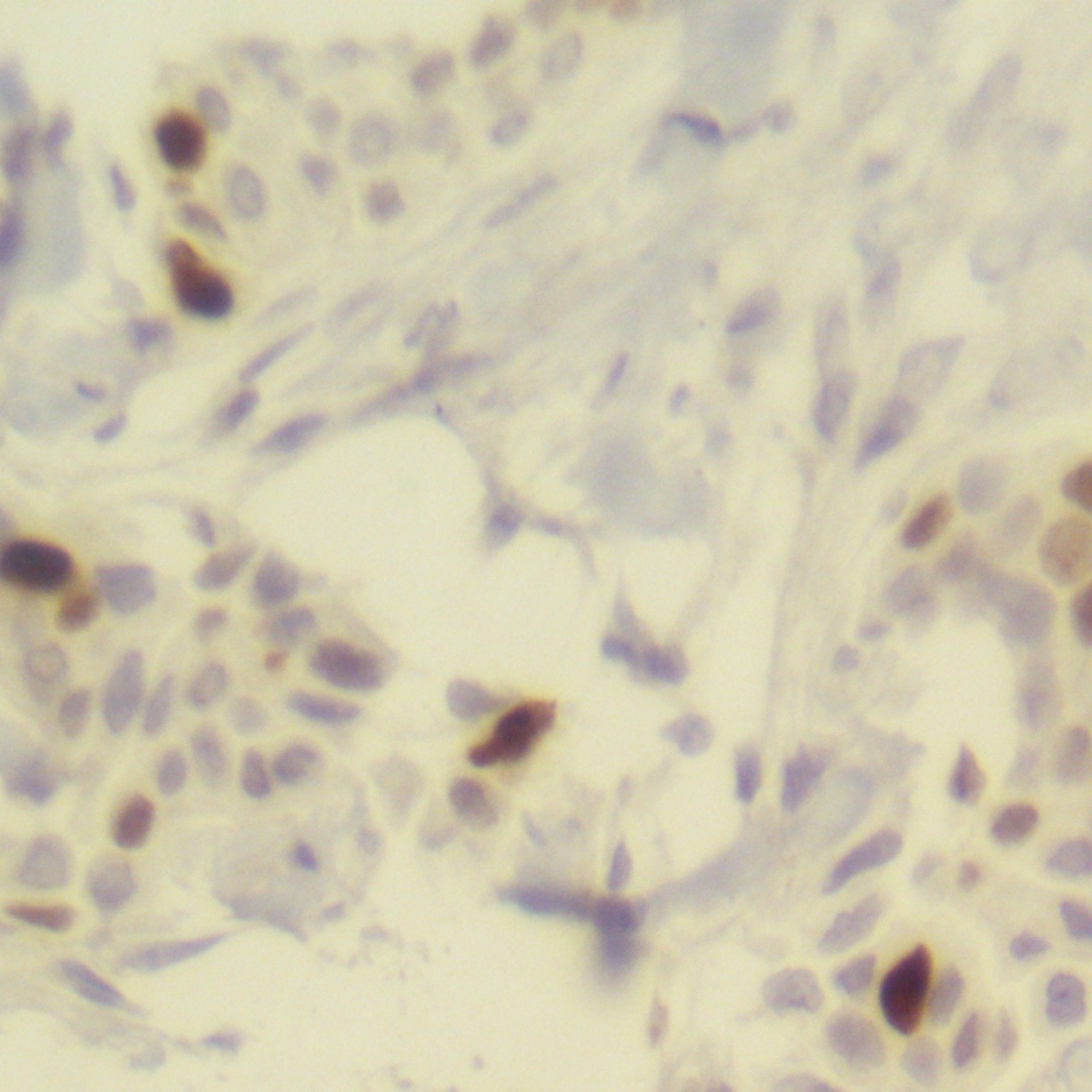}
        \caption{Original Tile}
        \label{fig:p12orig}
    \end{subfigure}
    \hfill
    \begin{subfigure}[b]{0.3\textwidth}
        \centering
        \includegraphics[width=\textwidth]{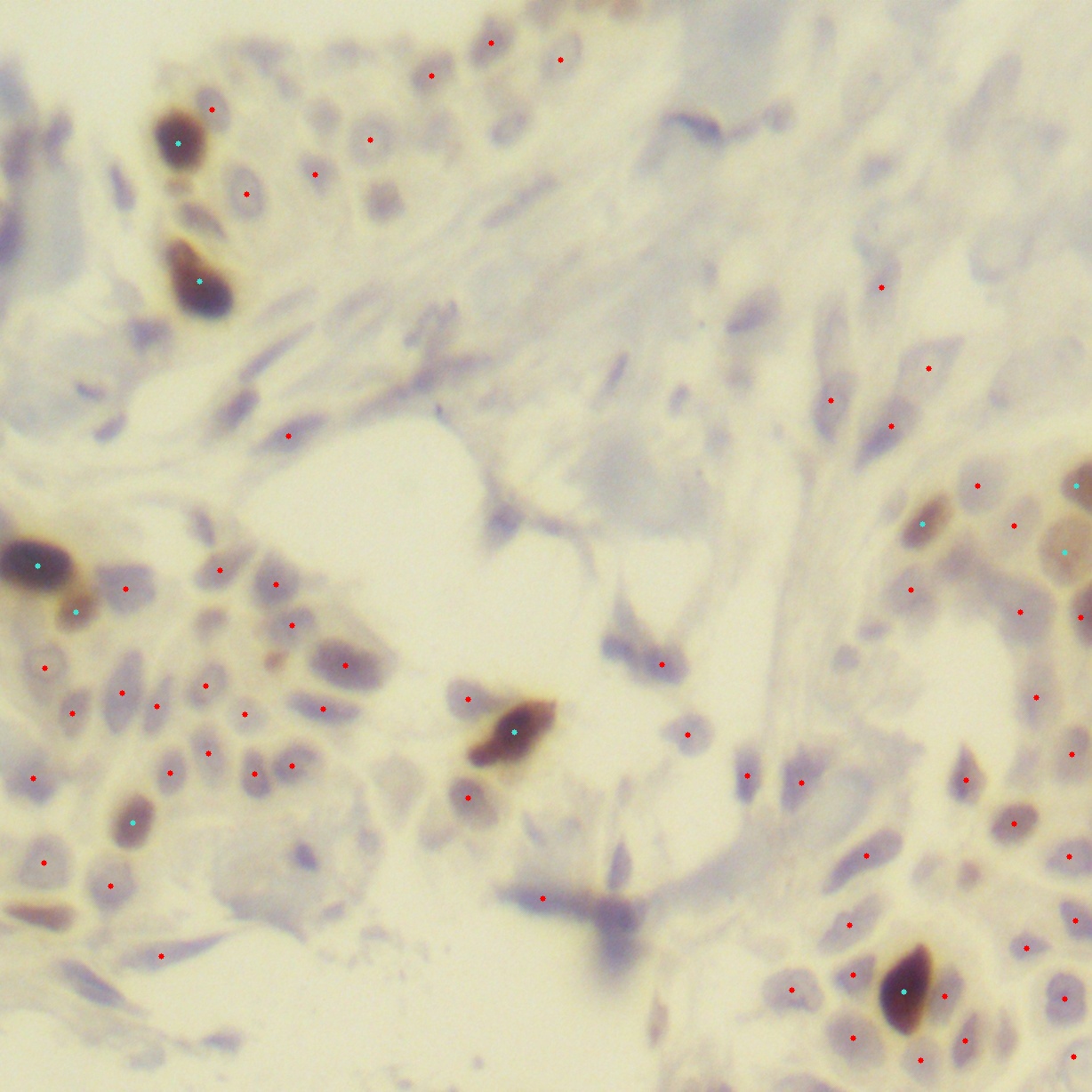}
        \caption{Keypoints}
        \label{fig:p12kp}
    \end{subfigure}
    \hfill
    \begin{subfigure}[b]{0.3\textwidth}
        \centering
        \includegraphics[width=\textwidth]{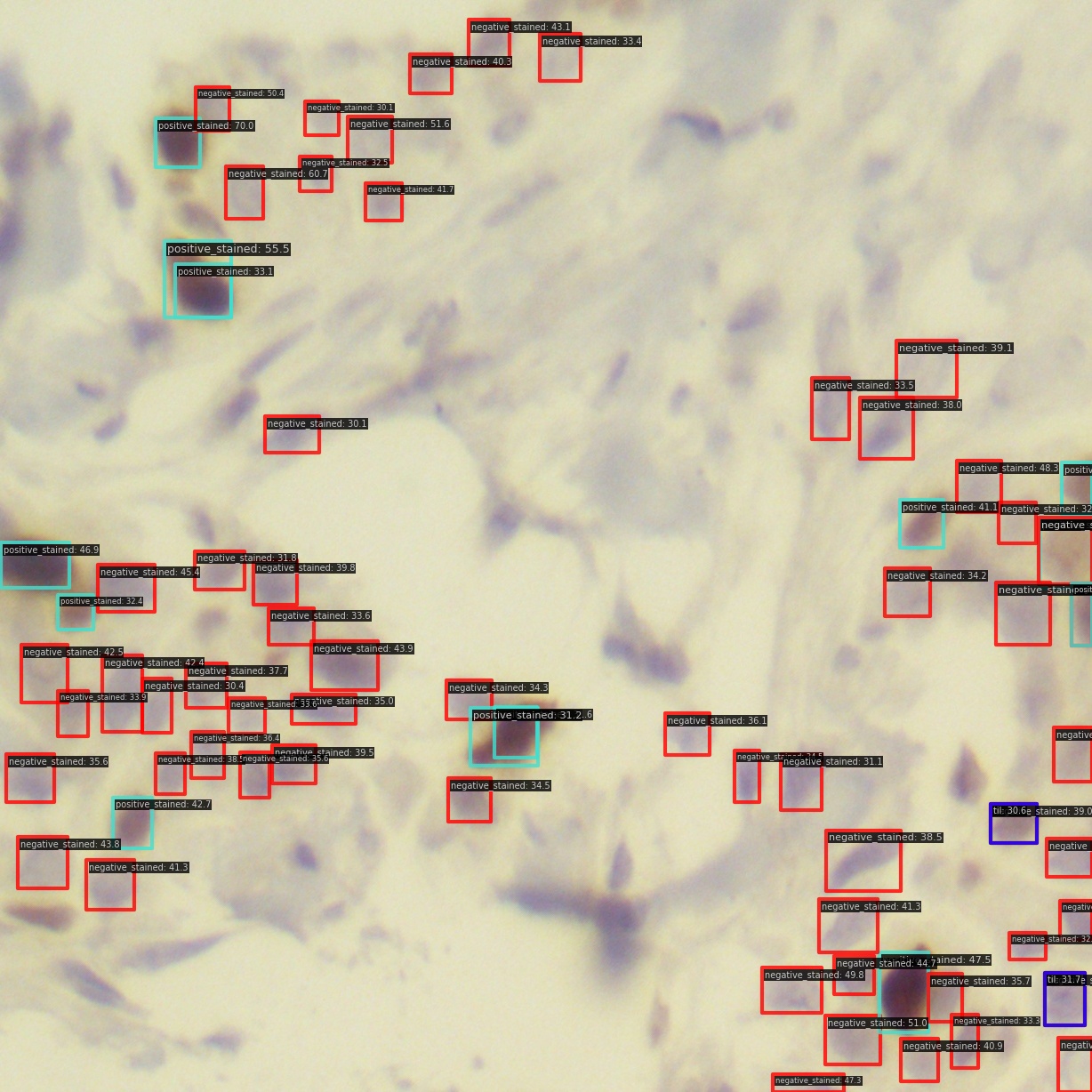}
        \caption{Predictions}
        \label{fig:p12bbox}
    \end{subfigure}
    \caption{Extracted tiles (\ref{fig:p12orig}) are annotated by domain expert with keypoints (\ref{fig:p12kp}) as ground truth. Once the cells of interest are labeled, the pre-trained NuClick neural network is applied to produce bounding box annotations to train an object detector (\ref{fig:p12bbox}).}
    \label{fig:p12_0168_2_gtvpred}
\end{figure}

The proliferation biomarker KI-67 \cite{scholzen2000ki,uxa2021ki,sun2018ki} is widely used in cancer diagnosis and is a valuable predictor of tumor aggressiveness, which is used to assess treatment options. Traditional manual methods of analyzing KI-67 staining are subjective and time-consuming. Additionally, manual estimation often exhibits inter and intra-observer variability \cite{karimi2020deep}. Accurately identifying the extent of KI-67 proliferation and tumor-infiltrated lymphocytes can help clinicians make informed decisions regarding the treatment of cancer patients. In recent years, deep learning \cite{lecun2015deep} based methods have been proposed to overcome these limitations. However, these methods often require pixel-level annotations on a cellular level, which can be expensive and time-consuming to obtain across whole slide pathology images. With improved accuracy, clinicians can obtain more accurate measurements, leading to more effective treatment plans, and ultimately improving the clinical outcomes of patients. Moreover, an automated estimation can save time and resources, allowing for more efficient, widespread use in clinical practices.

Deep neural networks are effective in a wide array of computer vision tasks, such as object detection and image classification \cite{voulodimos2018deep}. Recently, PathoNet \cite{negahbani2021pathonet} proposed a keypoint-based method for KI-67 detection, which achieved competitive performance compared to other deep learning-based methods with a more efficient keypoint annotation method. In parallel, RTMDet \cite{lyu2022rtmdet} was introduced as an efficient real-time object detector that outperforms the YOLO series \cite{DBLP:journals/corr/RedmonDGF15} and is easily extensible for various object recognition tasks, such as instance segmentation and rotated object detection.

In this work, we present a novel deep learning model, PathRTM, which builds upon RTMDet and extends the single pixel keypoint labels of the original PathoNet model to bounding box labels with real-time object recognition. PathRTM can be used to improve KI-67 proliferation and tumor-infiltrated lymphocyte estimation, as it can accurately detect KI-67 immunopositive, immunonegative, and lymphocyte cells. PathRTM has the potential to increase efficiency in digital pathology, as it can provide a more accurate and efficient KI-67 proliferation assessment.

Our work extends the original PathoNet method by incorporating bounding box-based supervision to improve performance. Specifically, we use the NuClick \cite{DBLP:journals/corr/abs-2005-14511} pre-trained neural network to convert the keypoint annotations into bounding box labels automatically, which are then used to train an RTMDet detector for efficient runtime while training and creating predictions when deployed. By maintaining the keypoints annotation, we maintain efficiency in the labeling process, single click per cell, while benefiting from higher-level supervision of the bounding boxes automatically.

\begin{figure}
	\centering
	\includegraphics[width=5cm]{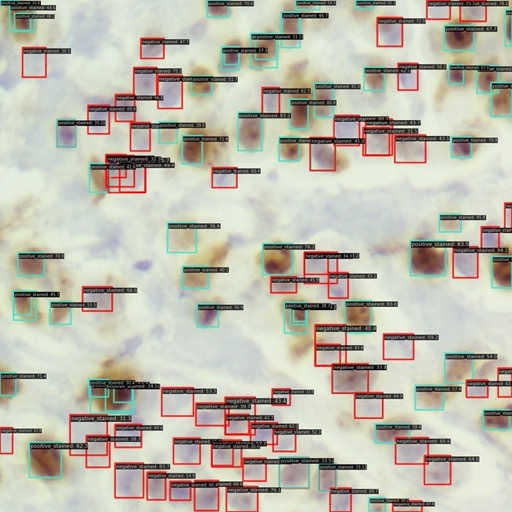}
	\hfill
	\includegraphics[width=5cm]{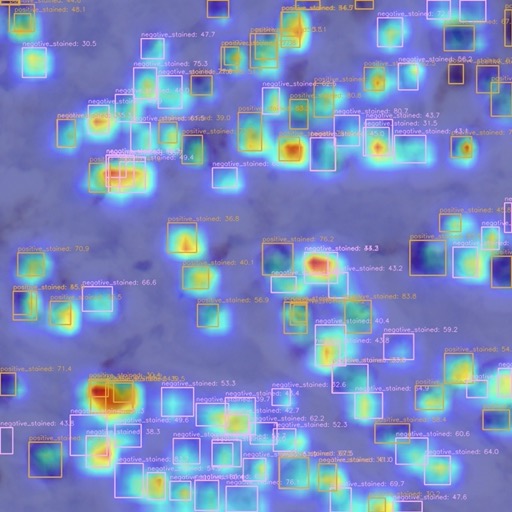}
	\caption{Visualization of PathRTM's predicted bounding boxes and GradCAM \cite{selvaraju1610grad} on a sample tile after training where the model displays its ability to detect the entire cell of interest apposed to a single keypoint within the cell.}
    \label{fig:gradcam}
\end{figure}

\section{Background}

\subsection{Deep Learning and Digital Pathology}

Following the introduction of AlexNet17 \cite{hinton2012imagenet}, which won the ImageNet challenge \cite{5206848} competition using a Convolutional Neural Network (CNN) \cite{lecun1995convolutional}, various research efforts have been conducted to apply deep learning to digital pathology \cite{madabhushi2016image}. Such efforts include applying deep learning in the multiple instance learning (MIL) framework for colon cancer classification \cite{xu2014deep}, automatic cell counting \cite{xie2018microscopy}, and deep learning methods for feature learning in breast cancer evaluation \cite{spanhol2016breast}.

\subsection{KI-67 Proliferation and Tumor-infiltrated Lymphocyte Estimation}
Previous works on KI-67 automation include Immunoratio \cite{tuominen2010immunoratio} and PathoNet \cite{negahbani2021pathonet}. Immunoratio is a method that differentiates between immunopositive and immunonegative cells to achieve automatic KI-67 scoring. Immunoratio relies on color deconvolution of images to segment on a pixel-level nucleus vs non-nucleus regions. Therefore dark artifacts, such as tissue folds, are likely to be misclassified. PathoNet is a U-Net \cite{DBLP:journals/corr/RonnebergerFB15} based deep network that uses single pixel keypoints for the detection of KI-67 immunopositive, immunonegative, and lymphocyte cells. In the final phase of predicting segmentation masks the PathoNet method applies the Watershed \cite{kornilov2018overview} algorithm to predict cell centers.

\subsection{NuClick}
While previous works apply deep learning for KI-67 automation, the availability of quality labeled datasets for KI-67 estimation is limited. PathoNet released a dataset with labels in the form of single pixel keypoints, which capture a single point of the cell. To extend PathoNet, we use NuClick, an efficient annotation tool, to generate bounding box labels from the keypoints in the original PathoNet dataset. Collecting labeled data is expensive and often requires domain expertise when applying DL in the medical imaging domain. NuClick demonstrates that one click inside each cell is enough to yield a precise annotation. The extension of keypoints to bounding boxes is beneficial from both a DL perspective and from a clinical perspecive providing clinicians with estimates of the sizes of the cells of interest from the bounding box predictions which often have clinical significance \cite{schob2018whole}.

\subsection{Real-time Multi-object Detection}
RTMDet is a real-time multi-object detection framework that builds upon the popular YOLO (You Only Look Once) architecture. RTMDet is designed to detect multiple objects within a scene with a high level of accuracy and speed. U-Net \cite{DBLP:journals/corr/RonnebergerFB15}, on the other hand, is a deep learning architecture commonly used for image segmentation tasks, which involves identifying and delineating specific objects or regions within an image. It consists of an encoder path that gradually reduces the spatial resolution of the input image and a decoder path that gradually recovers the original resolution while performing segmentation. The PathoNet model utilizes the U-Net deep neural network architecture for the segmentation of keypoints of pathology images because of its high accuracy. However, U-Net is slower in runtime in comparison to RTMDet due to its deep and complex architecture with a large number of convolutional layers.  U-Net's estimated runtime on an RTX3090 GPU is between 5ms to 10ms per image  while on a T4 GPU between 15ms to 30ms. By increasing the degree of supervision from one keypoint pixel per cell instance to the entire cell region, our work PathRTM achieves state-of-the-art performance with improved runtime when deploying RTMDet based architecture. 

\section{PathRTM Methodology}

\begin{figure}
	\centering
	\includegraphics[width=10cm]{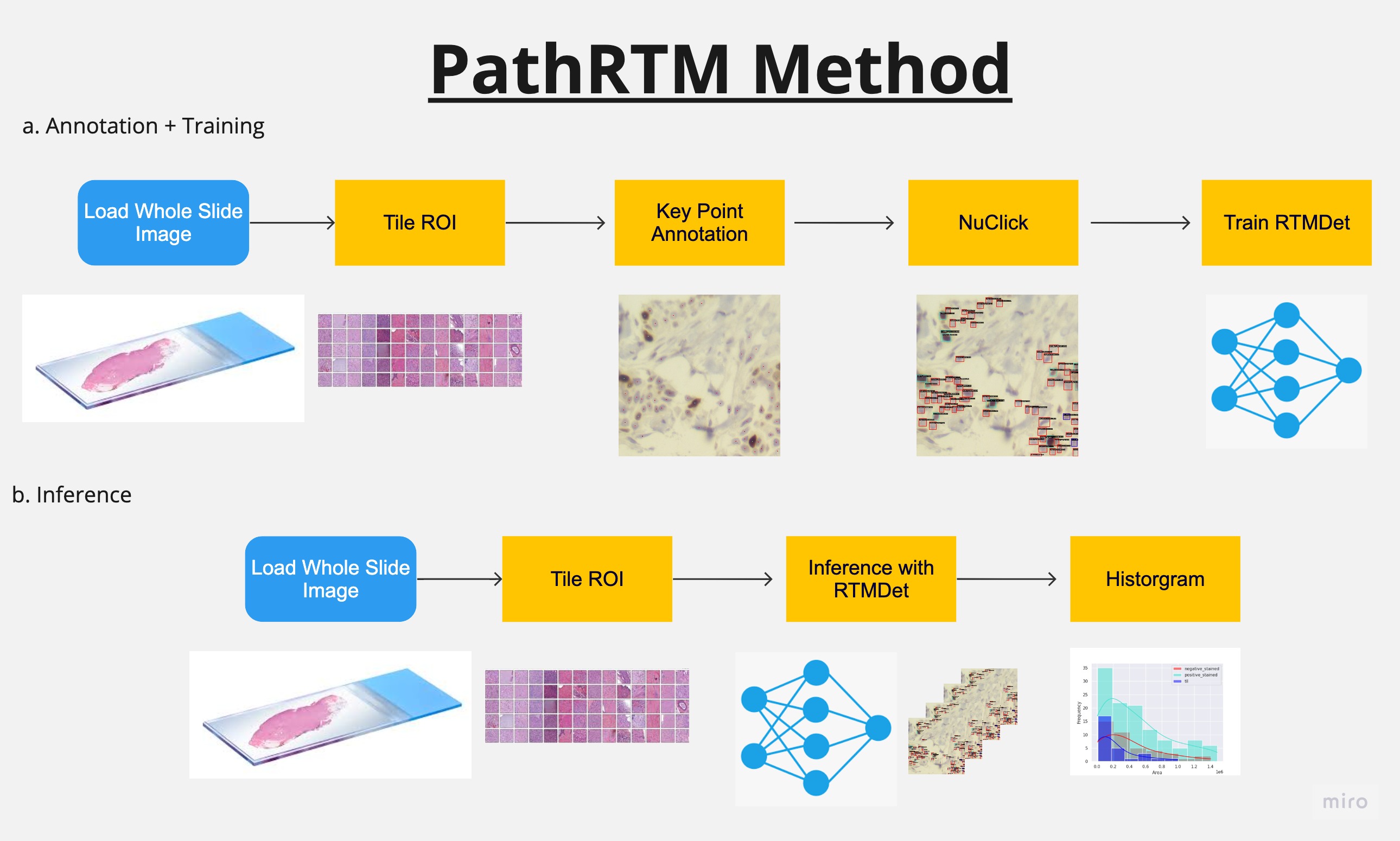}
	\caption{A diagram depicting the deployment of the PathRTM method. Step (a) the annotation of keypoints and automatic NuClick bounding box labels are generated, followed by training RTMDet detector. Step (b) the trained detector is deployed on a per-tile bases, the predicted bounding boxes over the tiles are grouped by patient identifiers, followed by cell level histograms over each patients predictions are generated. } \label{flow}
\end{figure}

An overview of the PathRTM method is depicted in Figure \ref{flow} and is described in detail below.

\subsection{ROI Tiling}

Whole slide pathology images are very large and complex, often containing a lot of information that can be difficult to interpret or analyze. In addition, whole slide images may contain irrelevant information such as marker outlining.  One way to make whole slide images more manageable is to split them into smaller tiles \cite{dimitriou2019deep}. In this particular case, the whole slide pathology images are being split into tiles of size 1228 by 1228 in the region of interest. This means that only the most relevant part of the image is being analyzed, rather than the entire image. This assists in reducing the computational load making the analysis more efficient. Additionally, by focusing only on the region of interest, we assist the detector in identifying important features or patterns that would otherwise be obscured by irrelevant or extraneous information.

\subsection{Keypoint Labeling}

When analyzing histopathology images, pathologists are interested in identifying specific cell types and their characteristics within the tissue sample. To accomplish this, pathologists manually annotate individual cells within the image. In this case, the pathologists are annotating by clicking on specific points, or keypoints, within the image to identify cells that are either immunopositive cells, immunonegative cells or tumor-infiltrating lymphocytes. This annotation is done on a per-tile basis, meaning for each whole slide each 1228 by 1228 tile is annotated individually.

\subsection{Automatic Keypoint to Cell Bounding Box}

After the pathologists have annotated the cells with immunopositive cells, immunonegative cells, and tumor-infiltrating lymphocytes in each tile, the next step is to apply a pre-trained NuClick neural network to these annotations (\ref{fig:p19_0121_1_gtbbox}). The purpose of applying this network is to convert each keypoint within each cell into a bounding box surrounding the cell. By casting each keypoint in a cell to a bounding box label, our method better understands each cell's features and surroundings. The NuClick neural network has been trained on a large dataset of annotated images to recognize and segment cells accurately.

\subsection{Training and Evaluation}

After automatically generating the higher level of supervision bounding box labels, we use RTMDet to train a model to detect tumor cells, immunopositive cells, immunonegative cells, and tumor-infiltrated lymphocytes in pathology images. To train the network, we used bounding box labels that were automatically generated. By training RTMDet on these labels, the network learns to recognize and classify each cell based on its featured characteristics (\ref{fig:gradcam}). We train the RTMDet detector for 1000 epochs. After training, we evaluated the performance of the detector by measuring its accuracy in detecting and classifying each cell from new images from the test set. This allowed us to ensure that the network was able to generalize well and accurately detect tumor cells in a wide range of pathology images.

\section{Results}

\begin{figure}
	\centering
	\includegraphics[width=5cm]{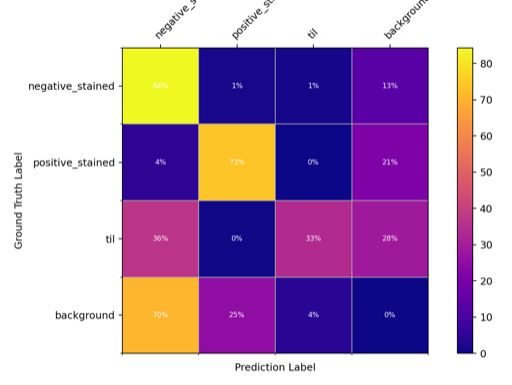}
	\caption{A confusion matrix with categories immunonegative, immunopositive, tumor-infiltrated lymphocytes, and background.  } \label{cm}
\end{figure}

The results of PathRTM, the RTMDet detector trained and evaluated on the generated bounding box labels from NuClick, are reported here. We used the specifically trained RTMDet-tiny model with image sizes of 640 by 640, 4.8 million parameters, and 8.1 GFLOPS. The runtime was measured in milliseconds for TRT-FP16 on RTX3090 and T4. RTX3090 achieves a latency of 0.98ms and the T4 achieves a latency of 2.34ms.

\begin{table}
	\centering
	\caption{PathRTM Overall AP results}\label{tab1}
	\begin{tabular}{|l|l|l|}
		\hline
		{\bfseries Model} & {\bfseries AP} & {\bfseries AP50} \\
		\hline
		Yolov8 nano       & 40.5\%         & 55.3\%           \\
		RTMDet tiny       & 41.3\%         & 56.7\%           \\
		RTMDet small      & 40.50\%         & 55.2\%           \\
		\hline
	\end{tabular}
\end{table}

We evaluated PathRTM on our extension of the PathoNet benchmark dataset using standard COCO \cite{lin2014microsoft} object detection evaluation metrics \cite{padilla2021comparative}. To execute the evaluation, we use the MMYOLO \cite{mmyolo2022} python library. PathRTM achieves an average precision (AP) of 41.3% and AP50 of 56.7%. For comparison, on the standard COCO dataset, RTMDet-tiny achieves AP of 41.0% and AP50 of 57.4%. These results indicate that PathRTM is able to achieve competitive performance on our dataset, even when compared to the general-purpose object detector RTMDet-tiny. Table~\ref{tab1} provides the overall AP, and Table~\ref{tab2} provides AP on a per-class basis, showing the performance of our method on different cell types.

Fig.~\ref{cm} is the confusion matrix after training. From this confusion matrix, we can observe that the PathRTM model is able to accurately differentiate between different cell types, with a majority of the cells being correctly classified. However, there are still some instances of misclassification, indicating potential areas for improvement in our model.

Upon analyzing the ground truth labels and the predicted labels, it seems often the case that several negative cells were not labeled in the original PathoNet dataset (see \ref{appendixlabeling}). This finding suggests that the original dataset may have some annotation inconsistencies, which could affect the overall performance of our method. Further work could focus on refining the dataset to ensure more accurate ground truth labels.

After training, we extract the cell pixel sizes histograms across all cells (Fig.~\ref{allhist}) and on a per-patient basis (see \ref{appendixhist}). The histograms provide insights into the distribution of cell sizes in our dataset, which can help us understand the characteristics of the cells being detected. By examining these histograms, we can gain a global overview of patients' overall prognosis by analyzing the distributions across different cell types and sizes, which can help us better understand the performance of our method in detecting cells with varying characteristics.

\begin{table}
	\centering
	\caption{PathRTM class AP results}\label{tab2}
	\begin{tabular}{|l|l|l|l|l|l|}
		\hline
		{\bfseries Model} & {\bfseries Immunonegative AP} & {\bfseries Immunopositive AP} & {\bfseries TIL AP} \\
        \hline
		Yolov8 nano       & 48.5\%                        & 46.7\%                        & 26.3\%             \\
        \hline
		RTMDet tiny       & 49.6\%                        & 48.0\%                        & 26.2\%             \\
		\hline
		RTMDet small       & 49.1\%                        & 47.3\%                        & 25.2\%             \\
        \hline
	\end{tabular}
\end{table}

\begin{figure}
	\centering
	\includegraphics[width=5cm]{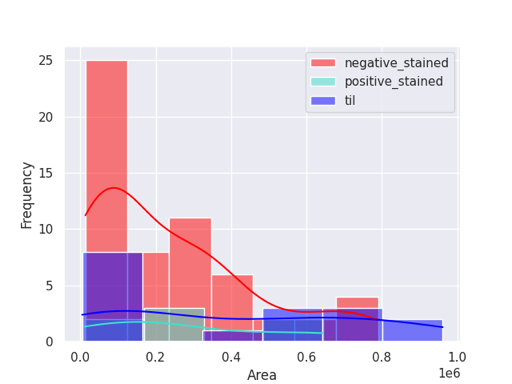}
	\caption{A histogram across all cells } \label{allhist}
\end{figure}

\FloatBarrier

\section{Conclusion}

In this paper, we propose PathRTM, an RTMDet-based deep neural network detector that leverages higher-level supervision to improve KI-67 proliferation and tumor-infiltrated lymphocyte estimation. Our experiments show that the proposed model achieves state-of-the-art KI-67 immunopositive, immunonegative, and lymphocyte bounding box detection. The proposed model can be used in the diagnosis and prognosis of cancer, which can have a significant impact on patient outcomes with our proposed method’s ability to estimate cells of interest sizes in realtime. Whole slide images are very large and detecting cells of interest is computationally expensive. Our proposed method can be used to detect cells of interest in realtime per tile, which can have a significant impact on the deployment of automatic pathology systems. Furthermore, our proposed method can be used to estimate the size of cells of interest. By providing a holistic view of the distribution of the cell types and sizes of the cells within an entire whole slide image, clinicians can better estimate the prognosis of a patient, the efficacy of treatment plans for a patient, and overall assist in providing a more accurate diagnosis.

%
% ---- Bibliography ----
%
% BibTeX users should specify bibliography style 'splncs04'.
% References will then be sorted and formatted in the correct style.
%
\medskip

\bibliographystyle{splncs04}
\bibliography{mybibliography}

\section{Appendix} \label{appendix}

\subsection{Labeling Analysis}\label{appendixlabeling}{
	Comparison of original tile, to ground truth keypoint annotation, to PathRTM predictions.

	\begin{figure}
		\centering
		\begin{subfigure}[b]{0.3\textwidth}
			\centering
			\includegraphics[width=\textwidth]{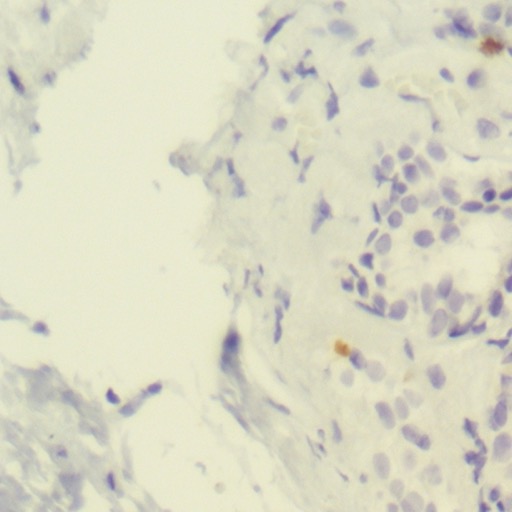}
			\caption{Original Tile}
			\label{fig:p19orig}
		\end{subfigure}
		\hfill
		\begin{subfigure}[b]{0.3\textwidth}
			\centering
			\includegraphics[width=\textwidth]{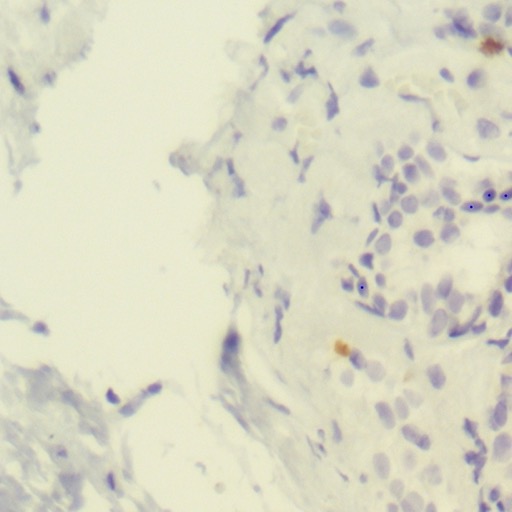}
			\caption{Keypoints}
			\label{fig:p19kp}
		\end{subfigure}
		\hfill
		\begin{subfigure}[b]{0.3\textwidth}
			\centering
			\includegraphics[width=\textwidth]{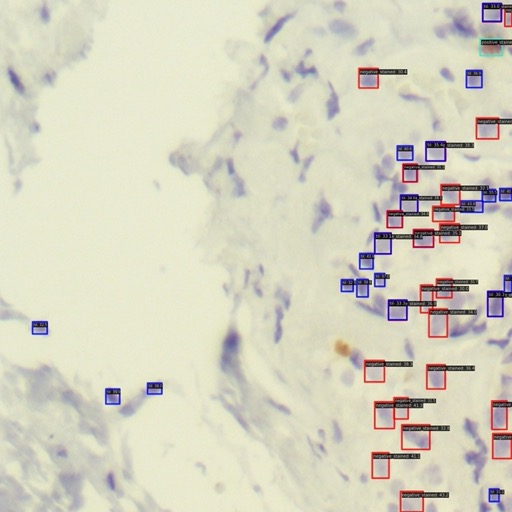}
			\caption{Predictions}
			\label{fig:p19bbox}
		\end{subfigure}
		\caption{p19 Original vs. Ground Truth vs. Prediction}
		\label{fig:p19_0121_1_gtbbox}
	\end{figure}

	\begin{figure}
		\centering
		\begin{subfigure}[b]{0.3\textwidth}
			\centering
			\includegraphics[width=\textwidth]{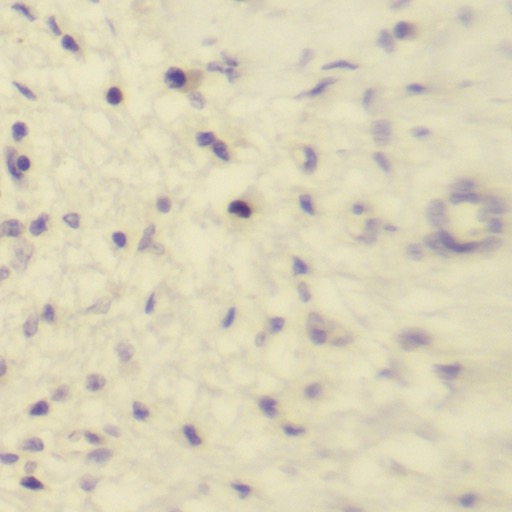}
			\caption{Original Tile}
			\label{fig:p1orig}
		\end{subfigure}
		\hfill
		\begin{subfigure}[b]{0.3\textwidth}
			\centering
			\includegraphics[width=\textwidth]{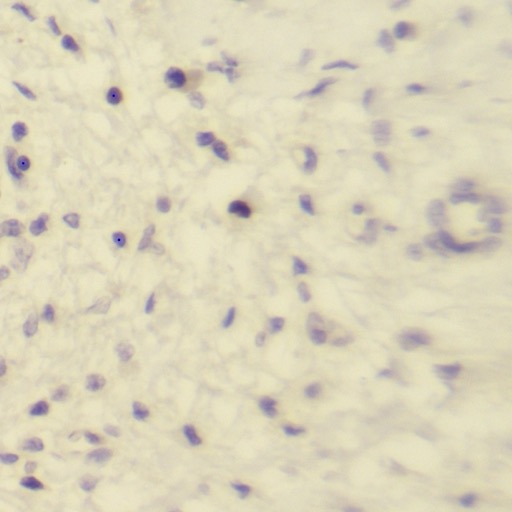}
			\caption{Keypoints}
			\label{fig:p1kp}
		\end{subfigure}
		\hfill
		\begin{subfigure}[b]{0.3\textwidth}
			\centering
			\includegraphics[width=\textwidth]{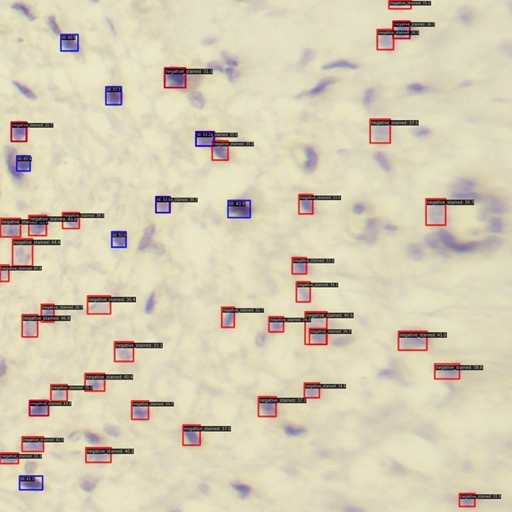}
			\caption{Predictions}
			\label{fig:p1bbox}
		\end{subfigure}
		\caption{p1 Original vs. Ground Truth vs. Prediction}
		\label{fig:p1_0310_8_gtvpred}
	\end{figure}
	    
}
    
\FloatBarrier
\subsection{Histograms}\label{appendixhist}{
	        
	\begin{figure}
		\centering
		\includegraphics[width=4cm]{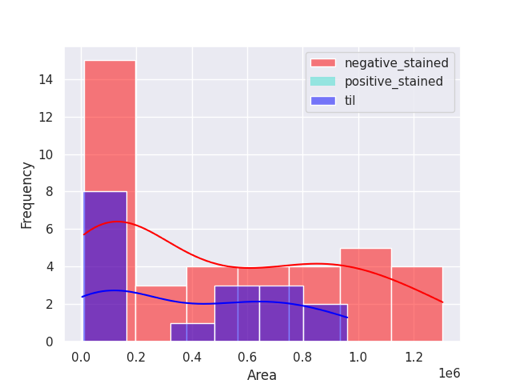}
		\caption{A histogram of p1 } \label{p1}
	\end{figure}
	        
	\begin{figure}
		\centering
		\includegraphics[width=4cm]{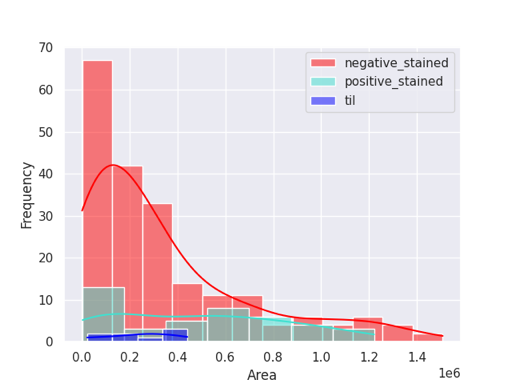}
		\caption{A histogram of p2 } \label{p2}
	\end{figure}

	\begin{figure}
		\centering
		\includegraphics[width=4cm]{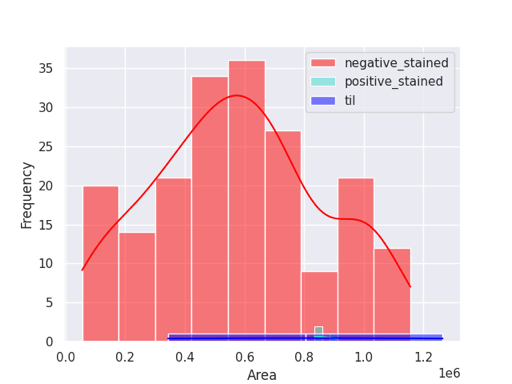}
		\caption{A histogram of p3 } \label{p3}
	\end{figure}

	\begin{figure}
		\centering
		\includegraphics[width=4cm]{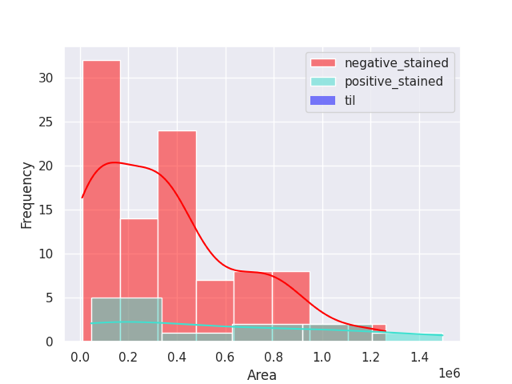}
		\caption{A histogram of p4 } \label{p4}
	\end{figure}
	Cell pixel sizes histograms per patient. 
}

\end{document}